\documentclass[prysrev]{revtex4-2}
\usepackage{hyperref}
\usepackage{textcase}
\usepackage{graphicx}
\usepackage{xcolor}
\usepackage{subcaption}
\usepackage{amsmath}
\usepackage{amssymb}
\usepackage{nicefrac, xfrac}
\usepackage{geometry}
\usepackage{multirow}
\usepackage{nccmath}
\usepackage[normalem]{ulem}	
\usepackage{natbib}

\begin{document}

\title{Acceleration of enzymatic reactions due to nearby inactive binding sites}

\author{Hila Katznelson}
\email{hilaweissman@campus.technion.ac.il}
\affiliation{Schulich Faculty of Chemistry, Technion-Israel Institute of Technology, Haifa 3200003, Israel}

\author{Saar Rahav}
\email{rahavs@technion.ac.il}
\affiliation{Schulich Faculty of Chemistry, Technion-Israel Institute of Technology, Haifa 3200003, Israel}

\begin{abstract}	
Many biological molecular motors and machines are driven by chemical reactions that occur in specific catalytic sites. We study whether the arrival of molecules to such an active site can be accelerated by the presence of a nearby inactive site. Our approach is based on comparing the steady-state current in simple models to reference models without an inactive site. We identify two parameter regimes in which the reaction is accelerated. We then find the transition rates that maximize this acceleration, and use them to determine the underlying mechanisms in each region.
In the first regime, the inactive site stores a molecule in order to release it following a reaction, when the neighboring catalytic site is empty. In the second regime, the inactive site releases a molecule when the catalytic site is full, in order to impede the molecules from leaving the active site before they react. For the storage mechanism, which is more likely to be biologically relevant, the acceleration can reach up to 15\%, depending on parameters.
\end{abstract}

\maketitle

	\section{Introduction}	
	\label{sec:intro}

There is an element of mutual enrichment between the study of living systems and the theory of nonequilibrium statistical mechanics. After all, life is composed of a host of out-of-equilibrium processes. Biological and biophysical systems often reveal new and interesting nonequilibrium phenomena. Research of such processes simultaneously expands the theory of nonequilibrium systems and deepens our understanding of living systems. 

A notable example of research
 on a biological process that advanced the understanding of out-of-equilibrium processes is the work of Berg, Winter, and Von Hipple \cite{berg1981diffusion}. They investigated the process by which a protein locates a target site on a DNA strand. Their conclusion was that a search mechanism combining free 3D diffusion with 1D diffusion along the DNA strand is more efficient than alternative mechanisms. Additional research improved our understanding of this mechanism \cite{slutsky2004kinetics,halford2004site}. These works have sparked interest in other aspects of DNA target site search processes \cite{gerland2002physical,shvets2018mechanisms,martin2015dynamics,alex2013speed}. A qualitatively similar but different search process is one where molecules arrive at an active site on the surface of a cell or a protein. This process also involves reduced dimensionality, but this time the molecules diffuse in a solution, reach a surface and then diffuse on the surface towards the active site \cite{berg1985diffusion,shin2018surface,misiura2021surface}.

Another example of a biological phenomenon with an intriguing out-of-equilibrium interpretation is cellular sensing. Cells can sense their environment through receptors on their surface, and the information acquired by these receptors enables bacteria to find nutrient-rich regions. Recent theoretical work studied the associated learning mechanisms, the energetic cost of sensing and maintaining accuracy, as well as the possible role of sensing with memory \cite{berg1977physics,hartich2016sensory,malaguti2021theory,lan2012energy,mehta2012energetic,sartori2014thermodynamic,govern2014energy}. The study of sensing demonstrates that many biological processes have information theoretical aspects. This is even more evident in processes that include the copying or transcribing of DNA or RNA, which are central to life. Several important works were devoted to study the nonequilibrium kinetics and thermodynamics of copying \cite{Hopefield1974,Ninio1975,Bennett1982}. Recent advances in our understanding of the thermodynamics of information \cite{Parrondo2015} have led to renewed interest in this problem \cite{Andrieux2008,Sartori2013,Bogod2017,Poulton_2021,Berx_2024}.

An interesting effect related to the arrival of molecules to catalytic sites was recently found by Zananiri {\it  et. al.} \cite{zananiri2022auxiliary}. They studied a bacterial helicase known as RecBCD, which exhibits an unusually fast unwinding rate, implying fast ATP hydrolysis. Based on a series of experiments, they deduced that RecBCD has several inactive binding sites for ATP, in addition to the two already known catalytic sites. After analyzing the biochemical and kinetic properties of these inactive sites, they concluded that these sites are necessary for achieving the fast unwinding rate. Another case of a biological mechanism involving inactive sites was recently presented by Mansson {\it et al } \cite{moretto2022multistep}. In their study, secondary Pi binding sites on myosin were shown to allow for gradual Pi release, thereby breaking the tight coupling between the mechanical and chemical steps in the cycle.

Zananiri {\it et. al.}'s work  focused specifically on RecBCD, but their results hint at a general out-of-equilibrium phenomenon worth studying on its own merits. How can an inactive binding site increase the arrival rate of molecules to a nearby active site? In this work, we address this question by building simple models and comparing them to nearly identical reference models that have no auxiliary sites. (In the following we used the terms auxiliary and inactive interchangeably to refer to a site that can capture and release molecules but do not catalyze a chemical process.) We allow the kinetics of the auxiliary site to depend on the state of the system, thereby introducing some allostery. While comparing the steady state reaction rates of both models, we make an effort to ensure that this comparison is fair. We identify two mechanisms of reaction acceleration. In one, the auxiliary site stores fuel molecules and releases them when the active site is empty. In the second mechanism, the auxiliary site releases a molecule to block the exit from the active site. This reduces the likelihood that a molecule in the active site escapes without undergoing catalysis. Each mechanism is dominant in a different parameter region.

This paper is organized as follows. In Sec. \ref{sec:gen}, we define some ground rules and constraints to ensure that the question we study is both well-defined and nontrivial. Using these guidelines, we construct a simple kinetic model that captures the effect in Sec. \ref{sec:model}. We also define two figures of merit that quantify the increase in reaction rate. In Sec. \ref{sec:regions}, we identify two parameter regions where the auxiliary site increases the steady state reaction rate. The transition rates that maximize the figures of merit in each regime are found in Sec. \ref{sec:max}, using a combination of analytical and numerical methods. In Sec. \ref{sec:mechanism}, we discuss the mechanisms of acceleration. A generalized model is defined and analyzed in Sec. \ref{sec:nearby}, reinforcing the physical interpretations of the mechanisms. We conclude in Sec. \ref{sec:conc}.

	\section{Underlying assumptions}
	\label{sec:gen}

Our goal is to find out how the inclusion of inactive binding sites can affect the turnover rate in models of out-of-equilibrium molecular motors and machines. We tackle this question by comparing models with and without such sites, which are named auxiliary sites. We are interested in enzymatic processes that are often encountered in biological systems. In such setups, energy-rich molecules are present in a well-mixed solution that serves as a molecule reservoir. These molecules find their way to a catalytic site, where they undergo a chemical reaction that drives some useful processes. The system then reaches a nonequilibrium steady state where the fuel molecules flow towards the active site and are consumed there. 

We start by making several physical assumptions that are designed to create a meaningful and fair comparison between models. Without such assumptions, drawing useful conclusions from the comparison becomes challenging, as it may refer to completely different models. First, we note that the setups we consider are open out-of-equilibrium systems, with a fluctuating number of fuel molecules. This many-particle aspect of the problem is important. It enables auxiliary sites to influence the turnover rate by storing fuel molecules during parts of the dynamics, and releasing them following certain events. Second, we restrict our studies to models for which the addition of the auxiliary site does not change the reaction rate in a trivial way. This means that the auxiliary site shall not be directly connected to the active site or modify the rate of the chemical reaction in it. Furthermore, the presence of an auxiliary site should not allow additional pathways for fuel molecules to enter the system from the molecule reservoir. (For instance, by creating a channel through the protein.) Otherwise, the additional path would also trivially enhance the rate of arrival of molecules to the catalytic site. In the following we satisfy these restrictions by connecting the auxiliary site only to a region that is adjacent to the active site.

The restrictions mentioned above are needed but may not be sufficient to ensure a fair comparison between models with and without auxiliary sites. Specifically, if we supplement the addition of the auxiliary site with a change of parameters that would by itself result in an increased reaction rate, then it
would not be justified to attribute the faster rate to the presence of the auxiliary site. An example for a problematic addition would be a modification of the rate of arrival to the catalytic site.
To avoid such confounding factors we restrict our investigation to models where the addition of auxiliary sites does not affect the rates of already existing transitions. 

This last restriction still allows nontrivial situations in which the rates of entering and leaving the auxiliary site depend on
the state of the system. Here we focus on rates that depend on the state of the active site. Such a dependence can be mediated by
elastic deformations of the protein. It will be demonstrated later that such a correlation between the auxiliary and catalytic sites can indeed result in increased reaction rates.

	\section{A simple model}
	 \label{sec:model}

\begin{figure}
	\centering
	\begin{subfigure}[b]{0.4\textwidth}
		\includegraphics[scale=0.25]{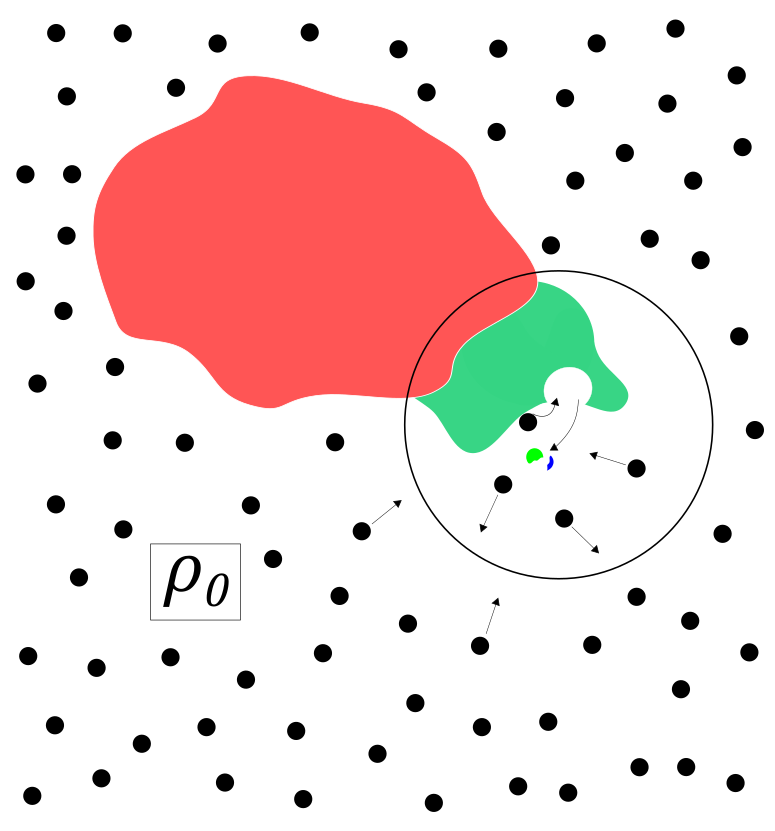}
		\caption{}
	\end{subfigure} 
	\begin{subfigure}[b]{0.4\textwidth}
		\includegraphics[scale=0.3]{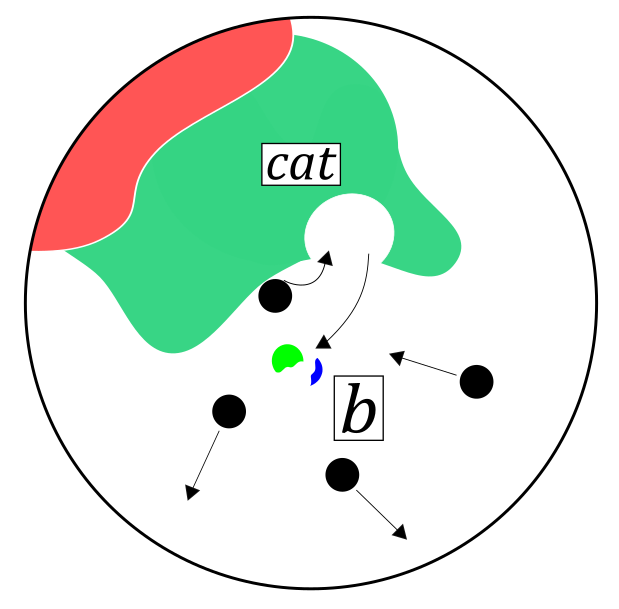}
		\caption{}
	\end{subfigure} 
	\begin{subfigure}[b]{0.4\textwidth}
		\includegraphics[scale=0.3]{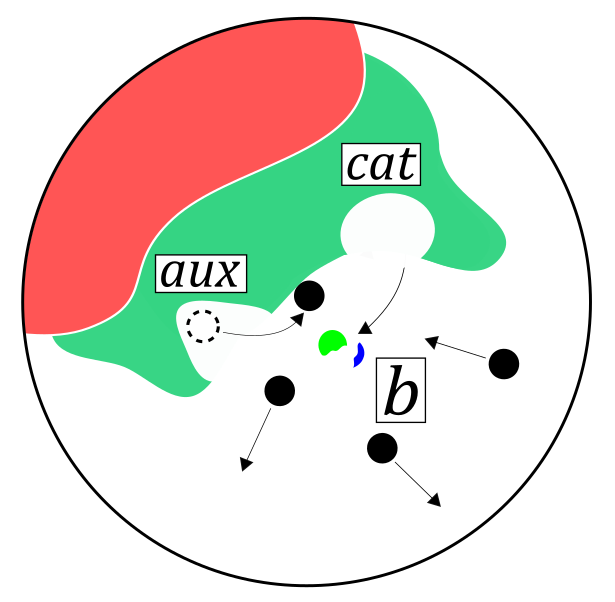}
		\caption{}
	\end{subfigure} 
	\caption{Heuristic representation of the setup of interest. (a) A protein functioning as a molecular machine or a motor is driven by a chemical reaction. Fuel molecules flow from a well-mixed environment to an active site, where they react. The environment functions as molecule reservoir of density $\rho_0$. (b) A close-up view of the region near the active site (the circled region in (a)). In our model, this region is coarse-grained into an effective site that bridges between other parts of the system. It is labeled by "b". The concentration of fuel molecules there may differ from that of the reservoir. The active site itself is labeled "cat". (c) A model that also has an auxiliary site, labeled "aux", in the vicinity of the catalytic site. A fuel molecule can bind to or leave this site but not undergo chemical reactions there. }
	\label{fig:protein}
\end{figure}

In this section, we construct the simplest model that qualitatively describes the possible role of auxiliary sites, while simultaneously satisfying the restrictions discussed in the previous section. For mathematical simplicity, our model has discrete states, whose dynamics follow a master equation. These discrete states are obtained by coarse-graining four spatial zones. A heuristic depiction of the whole setup is shown in Fig. \ref{fig:protein}, and each zone is labeled for easy identification.

One zone is the far away solution, which is assumed to be kept in equilibrium with fuel molecule concentration of $\rho_0$. It does not exhibit any dynamical behavior in our model. Its effects appear via the transitions in and out of the system with the respective rates $k_{+}\rho_{0}$ and $k_-$. Another is the active, or catalytic, site. Molecules enter this site with a transition rate $r_+$, escape back with the rate $r_-$, or undergo a chemical reaction with a rate $k_r$, which we assume is completely irreversible. This {\it absolute irreversibility} simplifies the dynamical description at the expense of assigning a divergent entropy production for the reaction step.

A third zone is a spatial region near the active site (and later also adjacent to the auxiliary site). We include this nearby region since we wish to avoid situations where the auxiliary site is directly connected to the active site. This region bridges between the other components of the system, and is therefore termed the bridging site in what follows. While it is perhaps more accurate to describe it as a spatial region in which molecules can diffuse, we elect to coarse-grain this zone into a single site. This is an approximation, but one that greatly simplifies the analysis, while still capturing the main qualitative features of the problem.

The final zone is the auxiliary site. It is connected directly to the bridging site but, importantly, not to either the fuel molecule reservoir, or the active site. The rates of entering and leaving the auxiliary site depend on the occupation of the active site. When it is empty, the transition rates are $h_+$ and $h_-$ respectively, and when it is occupied the rates turn to $\tilde{h}_+$ and $\tilde{h}_-$. This can model a deformation of the protein due to elastic stresses that follow capture of a molecule in the active site. Conveniently, the steady state of a model without the auxiliary site can be described by taking the limit in which the rates of entering this auxiliary site vanish.

\begin{figure}[htb]
		\centering
		\begin{subfigure}{\linewidth}
			\includegraphics[scale=0.6]{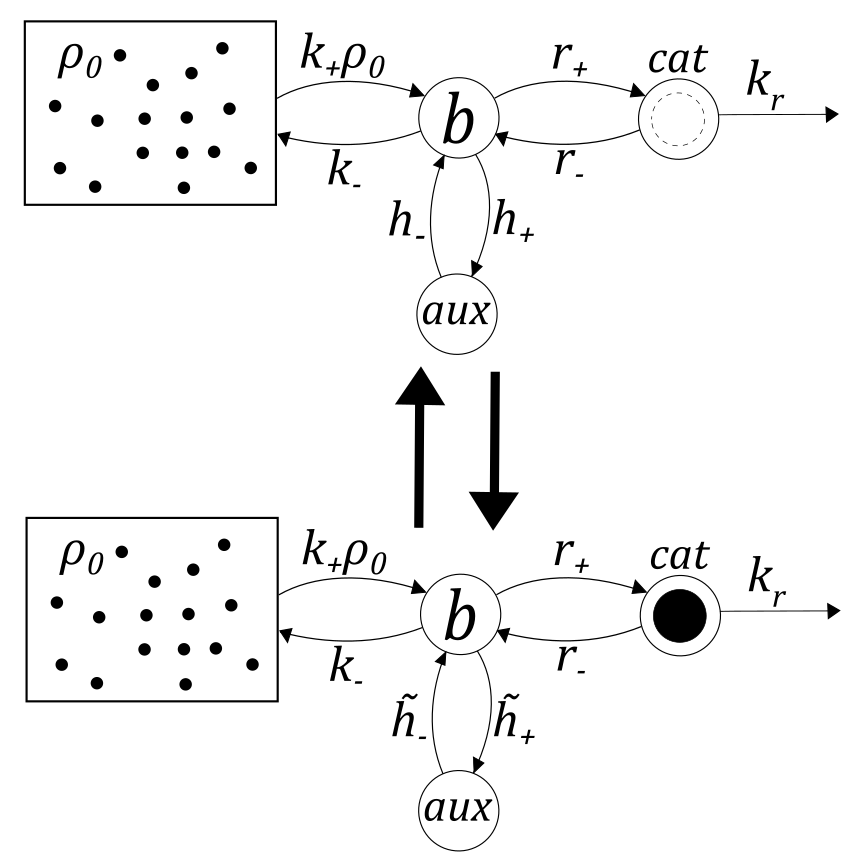}
			\label{fig:3sitescoop}
		\end{subfigure}
		\caption{A schematic depiction of the kinetics. The model has three sites: "cat" stands for the catalytic site, "aux" for the auxiliary site, and "b" for the bridging site. This last effective "site" connects the system to an infinite molecule reservoir with constant concentration $\rho_{0}$. The rate constants of the auxiliary site are $h_{+}$ and $h_{-}$ for an empty catalytic site, in the upper figure, and are $\tilde{h}_+$ and $\tilde{h}_-$ for a full catalytic site, in the lower figure.}	
		\label{fig:3sites}
	\end{figure}

The result of this coarse-graining is schematically depicted in Fig. \ref{fig:3sites}. All possible transitions between sites are represented by arrows, and the associated rates are given beside them. This schematic depiction is only meant to show how fuel molecules can flow from the reservoir to the active site. To complete the description, one must remember that several fuel molecules can be present simultaneously in the system. This is a many-particle problem, and should be described in terms of the relevant many-particle states and the transitions between them. Here we make the simple assumption that each of the regions can be either empty, or include one fuel molecule. Namely, we assume that fuel molecules block each other, and once a site has one, others can no longer enter it. While this assumption is reasonable for the active and auxiliary sites, it is clearly a crude approximation for the bridging site. Nevertheless, we first study a model with this assumption. Only later, in Sec. \ref{sec:nearby}, we examine how the results are modified when this assumption is relaxed.

\begin{figure}[htb]
		\centering
		\begin{subfigure}[b]{\textwidth}
				\includegraphics[scale=0.8]{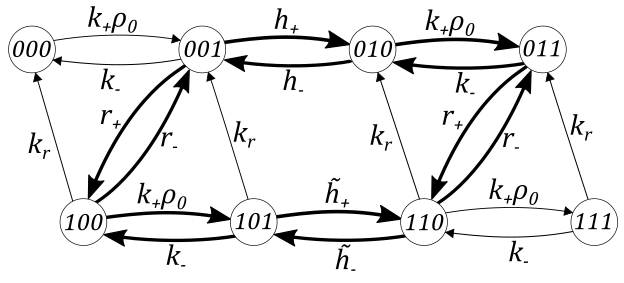}
				\caption{}
				\label{fig:8stscoopgraph}
		\end{subfigure}
	\begin{subfigure}[b]{0.8\textwidth}
		\includegraphics[scale=0.32]{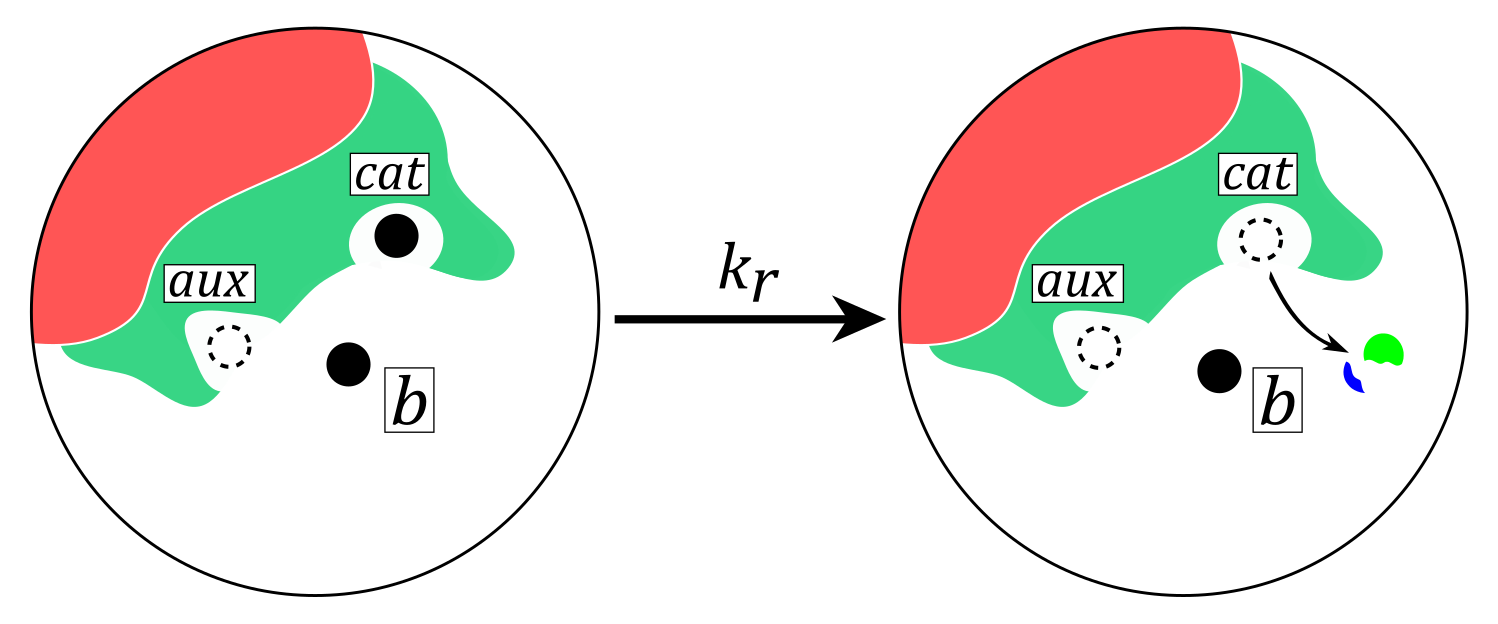}
		\caption{$\sigma=(101)\rightarrow\sigma=(001)$}
		\label{fig:examp101to001}
	\end{subfigure}

	\caption{(a) A graph representation of the master equation depicting the model. Nodes correspond to many-particle configurations of the system, ordered according to the occupation numbers  $n_c n_a n_b$. Arrows represent the physically allowed transitions, with the corresponding transition rate marked. Note that a physical process, such as the reaction at the active site, can appear more than once due to different arrangements of spectator molecules that are not involved in the transition. The loop in bold is the only closed cycle in the model that does not include the completely irreversible transition of the chemical reaction (with rate $k_r$). (b) An illustration of one of the processes in the graph. Starting from $(101)$, a state with full active and bridging sites, the particle in the active site can undergo a reaction, leaving the system in state $(001)$.}
	\label{fig:8stscoop}
	\end{figure}

It is convenient to identify the many-particle states of the model using the three occupation numbers, $n_c, n_a, n_b$, corresponding to the occupation of the catalytic, auxiliary, and bridging sites. For instance $\sigma=(101)$ corresponds to $n_c=1, n_a=0, n_b=1$, which is the state with full active and bridging sites and an empty auxiliary site. Our model has therefore eight many-particle states. The possible transitions between those states, taking into account that molecules block each other, are depicted in Fig. \ref{fig:8stscoopgraph}. Each node of the graph denotes a many-particle state, and each arrow represents an allowed transition, whose matching process is deducible from the many-particle states it connects. The transition rates are shown next to each arrow. An example is depicted in Fig. \ref{fig:examp101to001}.

The probabilities to find the system in these many-particle states evolve according to the master equation 
	\begin{equation}
		\frac{d\overrightarrow{P}_{(t)}}{dt}=R\overrightarrow{P}_{(t)}.
		\label{eq:mastereq}
	\end{equation}
Here, $\overrightarrow{P}_{(t)}=\left(P(000), P(001), \cdots, P(111) \right)^\top$ is the vector of probabilities for the many-particle states, and $R$ is the rate matrix. Its off-diagonal entries correspond to the transition rates denoted in the figure, while its diagonal entries are minus the escape rate from each state, ensuring probability conservation. An exact expression for $R$ is given in appendix \ref{app:Rmat}.

We are interested in the non-equilibrium steady state of the model, denoted by $\Pi (n_c n_a n_b)$. This steady state satisfies
    \begin{equation}
    	R\overrightarrow{\Pi}=0.
    \end{equation}
The physical quantity of interest is the rate of chemical reaction at steady state, given by the current
	\begin{equation}
		J_{cat}=k_{r}\Pi (n_c=1)=k_r\cdot \left[\Pi (100)+\Pi (110)+\Pi (101)+\Pi (111) \right],
		\label{eq:jcat}
	\end{equation}
where $\Pi (n_c=1)$ is the probability for a full catalytic site. It is the sum of the probabilities of all four many-particle states with an occupied catalytic site. Since the maximal occupation is $1$, the reaction rate satisfies $0\leq J_{cat}\leq k_r$.

Our goal is to find out when this chemical reaction is accelerated by the addition of the auxiliary site. As mentioned earlier, this can be studied by comparing the current to that in a corresponding, {\it reference} system, which is without an auxiliary site but otherwise identical. This reference current, $J_0$, can be obtained simply by substituting $h_{+}\textmd{, }\tilde{h}_+=0$ to the expression for the steady state current.

One can define various quantities that compare the currents $J_{cat}$ and $J_0$. In the following we focus on two figures of merit that seem to be naturally suited for this goal.
Specifically, we use
\begin{equation}
	\Delta J_{cat}=J_{cat}-J_{0},
	\label{eq:diffjcat}
\end{equation}
and
\begin{equation}
	\frac{\Delta J_{cat}}{J_0}=\frac{J_{cat}}{J_0}-1.
	\label{eq:ratiodiffjcat}
\end{equation}
We look for parameter regions where these measures are positive, and give physical interpretation to the mechanisms by which the auxiliary site accelerates the reaction. Studying more than one figure of merit is required because the two quantities emphasize different regions of parameter space. $\Delta J_{cat}$ allows us to focus on the largest absolute increase in the reaction rate. In contrast, $\Delta J_{cat}/J_0$ highlights large relative accelerations, which can be highly relevant in regions where both $J_{cat}$ and $J_0$ are very small.

\section{Identification of parameter regions with increased reaction rates}
\label{sec:regions}

The current difference $\Delta J_{cat}$ can be calculated analytically from the steady-state probabilities of many-particle states. The result is rather cumbersome, but can be
recast as 
\begin{equation}
\Delta J_{cat} = B r_+ \left( r_--r_+\right)\left[ h_{+} \tilde{h}_- - Q \tilde{h}_+ h_{-}\right],
\label{eq:deltajc}
\end{equation}
The expressions for $ B$ and $Q$ are given in Appendix \ref{app:coeffdelJ}. They are positive for any positive value of the transition rates.

The fact that ${B}\textmd{, }Q>0$ is crucial. As consequence, the sign of $\Delta J_{cat}$ is determined by the simple factor $\left( r_--r_+\right)\left[h_{+} \tilde{h}_-- Q \tilde{h}_+h_{-}\right]$. This enables us to identify two parameter regimes in which the auxiliary site can accelerate the reaction. In the first region of parameters, which we call the blocking regime, we have
\begin{eqnarray}
 r_- & >  &r_+, \nonumber \\
 Q \textmd{ }& < &\frac{h_{+} \tilde{h}_-}{\tilde{h}_+h_{-}}.
\label{eq:condblocking}
\end{eqnarray}

We term the second regime the molecule storage regime. Here
\begin{eqnarray}
 r_- & <& r_+ , \nonumber \\
Q \textmd{ }& > &\frac{h_{+} \tilde{h}_-}{\tilde{h}_+h_{-}}.
\label{eq:condstorage}
\end{eqnarray}
The names for the regimes will become clear later, when we examine how the rates determine the typical processes that occur in the steady state. Both regimes are also applicable for $\Delta J_{cat}/J_0$, as the denominator $J_0$ is positive and can not change the sign.

\section{Maximizing the figures of merit $\Delta J_{cat}$ and $\Delta J_{cat}/J_0$ }
\label{sec:max}

In each of the two regimes defined in the section above, we search for the transition rates that would result in maximal $\Delta J_{cat}$ or $\Delta J_{cat}/J_0$. To make the problem well-defined, we restrict all the rates to a finite range of values, e.g. $h_{min} \le h_i \le h_{max}$. Without such restriction maximization of a quantity often leads to nonphysical diverging transition rates. We also recall that the change in reaction current should not be affected directly by the flow of fuel molecules from the reservoir or the reaction at the active site. We therefore view $k_r$ and $k_+ \rho_0$ as fixed parameters. In the following we set $k_r=10$, which is just a way of choosing the typical time scale of our model. In contrast, we consider several different values of $k_+ \rho_0$, in order to represent different physical situations. This leaves seven transition rates, $h_\pm, \tilde{h}_\pm, r_\pm$, and $k_-$, that are allowed to vary in the range $[1,100]$. These maximal and minimal rates are selected so they can be larger or smaller than $k_r$. This optimization over seven parameters is nontrivial, and is done by first determining some of the transition rates analytically, and then the rest numerically.

\subsection{Analytical derivation for the optimal values of the auxiliary site rates}
\label{subsec:h}

The values of $h_{\pm} \textmd{ and }\tilde{h}_\pm$ that maximize $\Delta J_{cat}$ and $\Delta J_{cat}/J_0$ turn out to depend only on the sign of $r_--r_+$, without dependence on the values of the remaining rates. They can be found analytically due to relatively simple dependence of Eq. (\ref{eq:deltajc}) on these rates. A direct calculation of the partial derivative in e.g. $h_{+}$ leads to 
\begin{equation}
\frac{\partial \Delta J_{cat}}{\partial h_{+}} = {\cal M} r_+ \left( r_--r_+\right).
\label{eq:derivdelta}
\end{equation}
${\cal M}$, which is given in appendix \ref{app:coeffdelJ} , depends on all the rates. Importantly,  ${\cal M}>0$, so $\Delta J_{cat}$ is always monotonous in $h_+$, and its optimal value is either the maximal or minimal rate, depending on the sign of $r_--r_+$. We therefore conclude that $h^*_{+}=h_{max}$ in the blocking regime, and $h^*_{+}=h_{min}$ in the storage regime, irrespective of the values of the rest of the rates.

This calculation can be repeated for the rates $h_-$, $\tilde{h}_+$, and $\tilde{h}_-$. The calculation for $\tilde{h}_{-}$ is essentially the same as for $h_+$. However, the signs of the derivatives are opposite for
$h_{-}$ and $\tilde{h}_{+}$. The rest of the calculation proceeds according to the same lines. As a result, the optimal values of the auxiliary site rates are determined by the parameter regime, and are otherwise independent of the precise values of the rates. We find that
\begin{equation}
	 h^*_{+}=\tilde{h}^{*}_{-}=h_{max}, \text{ and }
	 h^*_{-}=\tilde{h}^{*}_{+}=h_{min},
\end{equation}
in the blocking regime. Similarly, we obtain
\begin{equation}
		h^*_{+}=\tilde{ h}^{*}_{-}=h_{min}, \text{ and }
			h^*_{-}=\tilde{h}_{+}=h_{max},
\end{equation}
in the storage regime. We can now substitute these values in each regime, and search for the maximum as a function of the remaining transition rates, $r_\pm$ and $k_-$.

\subsection{Numerical optimization}
\label{subsec:numopt}

The three remaining transition rates that maximize the two figures of merit are found using numerical optimization, performed with the help of Wolfram Mathematica and Matlab.  
In both programs, we consider the blocking and particle storage regimes separately for each figure of merit, namely, $\Delta J_{cat}$ and $\Delta J_{cat}/J_0$. Particularly, we substitute the respective optimal values of the auxiliary site rates, and optimize the remaining rates within each regime. 

The results for the maximization of $\Delta J_{cat}$ are presented in Tables \ref{table:1} and \ref{table:2}. We use $k_r=10$, and examine three different fuel molecules 
concentrations, $k_+ \rho_0 =1,10,100$. These values are chosen to demonstrate the behavior when fuel molecules are scarce, in abundance, or exhibit moderate concentrations.

\begin{table}[h]
	\centering
	\begin{subtable}{\linewidth}
		\centering
		\begin{tabular}{||c | c | c | c | c | c | c||} 
			 \hline
			$k_+ \rho_0$ & $ r_+^*$ & $ r_-^*$ & $k_-^*$ & $\Delta J_{cat}^*$  & $J_{cat}$ & $J_0$
			\\ [0.5ex] 
			\hline\hline
			1 & 6.87 & 100 & 1 & 0.294 & 0.546 & 0.251 \\ 
			10 & 14.4 & 100 & 1 & \textbf{1.19} & 2.78 & 1.59 \\
			100 & 13.3 & 100 & 3.93 & 0.498 & 4.14 & 3.64 \\ [1ex] 
			\hline
		\end{tabular}
		\caption{Blocking regime.}
		\label{table:1}
	\end{subtable}
\quad%
\begin{subtable}{\linewidth}
	\centering
	\begin{tabular}{||c | c | c | c | c | c | c||} 
		\hline
		$k_+ \rho_0$ & $ r_+^*$ & $ r_-^*$ & $k_-^*$ & $\Delta J_{cat}^*$  & $J_{cat}$ & $J_0$
		\\ [0.5ex] 
		\hline\hline
		1 & 100 & 1 & 13.5 & 0.0298 & 0.851 & 0.821 \\ 
		10 & 100 & 1 & 11.3 & \textbf{0.738} & 5.98 & 5.24 \\
		100 & 100 & 3.22 & 100 & 0.273 & 8.05 & 7.78 \\ [1ex] 
		\hline
	\end{tabular}
	\caption{Molecule storage regime.}
	\label{table:2}
\end{subtable}
	\caption{The two sets of transition rates that maximize $\Delta J_{cat}$, calculated for several values of $k_+ \rho_0$. The maximal value of $\Delta J_{cat}$ in each regime is highlighted using a bold font.}
	\label{tab:delJ}
\end{table}
In both regimes it seems that the auxiliary site has the largest effect when $k_+ \rho_0 \simeq k_r$. 
A possible explanation is the difficulty for the auxiliary site to substantially increase the reaction rate when the active site is nearly always occupied, or when the system is predominantly empty.
A comparison between the different regimes shows that larger current differences $\Delta J_{cat}$ exist in the blocking regimes. Also, for most cases we find that at least one
of the rates $r_-^*$ or $r_+^*$ is found at the boundary of the allowed values. However, contrary to the situation in Sec. \ref{subsec:h}, we have no analytical proof that this is always the case.

We turn to examine the rates that maximize the second figure of merit, $\Delta J_{cat}/J_0$. This choice is motivated by the possibility that a diminished fuel molecule concentration will result in low values of $\Delta J_{cat}$, primarily due to a small rate of arrival to the active site. Under those conditions the ratio $\Delta J_{cat}/J_0$
highlights the relative enhancement of the reaction, which need not be small. Tables \ref{table:3} and \ref{table:4} are similar to Tables \ref{table:1} and \ref{table:2}, but present
the rates that maximize $\Delta J_{cat}/J_0$. The most noticeable aspect of these results is the sizable current's enhancement in the blocking regime with $k_+ \rho_0=1$. There, the auxiliary site is found to increase the reaction rate by a factor of 2.89. This is a fairly strong acceleration.

\begin{table}[h]
	\centering
	\begin{subtable}{\linewidth}
		\centering
		\begin{tabular}{||c | c | c | c | c | c | c||} 
			 \hline
			$k_+ \rho_0$ & $ r_+^*$ & $ r_-^*$ & $k_-^*$ & $\left(\Delta J_{cat}/J_0\right)^*$  & $J_{cat}$ & $J_0$
			\\ [0.5ex] 
			\hline\hline
			1 &1 & 100 & 1.25 & \textbf{2.89} & 0.162 & 0.0418 \\ 
			10 & 1 & 100 & 1 & 1.87 & 0.409 & 0.143 \\
			100 & 1 & 100 & 15.9 & 0.293 & 0.355 & 0.274 \\ [1ex] 
			\hline
		\end{tabular}
		\caption{Blocking regime.}
		\label{table:3}
	\end{subtable}
    \quad%
	\begin{subtable}{\linewidth}
		\centering
		\begin{tabular}{||c | c | c | c | c | c | c||} 
			 \hline
			$k_+ \rho_0$ & $ r_+^*$ & $ r_-^*$ & $k_-^*$ & $\left(\Delta J_{cat}/J_0\right)^*$  & $J_{cat}$ & $J_0$
			\\ [0.5ex] 
			\hline\hline
			1 & 100 & 1 & 18.8 & 0.0372 & 0.808 & 0.779 \\ 
			10 & 100 & 1 & 21.8 & 0.147 & 5.46 & 4.76 \\
			100 & 100 & 13.5 & 100 & 0.0365 & 7.44 & 7.17 \\ [1ex] 
			\hline
		\end{tabular}
		\caption{Molecule storage regime.}
		\label{table:4}
	\end{subtable}
	\caption{The two sets of transition rates that maximize $\Delta J_{cat}/J_0$ for several values of $k_+ \rho_0$. The maximal value of  $\Delta J_{cat}/J_0$ in the blocking regime is highlighted using a bold font.}
	\label{tab:delJ/J0}
\end{table}

\section{Mechanisms of acceleration}
\label{sec:mechanism}

The numerical and analytical results presented in Sec .\ref{sec:max} suggest two qualitatively distinct mechanisms by which the auxiliary site can increase the reaction rate.
These can be understood by looking at the transition rates that maximize the figures of merit $\Delta J_{cat}$ and $\Delta J_{cat}/J_0$.

\subsection{Blocking mechanism}

The blocking regime is characterized by the rates $r^*_-=100 > r^*_+$, which then leads to $h_{+}^*=\tilde{h}_{-}^*=100$, and to $h_{-}^*=\tilde{h}_{+}^*=1$. The largest 
values of the figures of merit are found in this regime. Specifically, the greatest acceleration is found for $k_+ \rho_0 = 10$.
Under these conditions the auxiliary site can increase the reaction rate from $J_0 \simeq 1.59$ to $J_{cat} \simeq 2.78$.
An even more striking improvement is observed in the relative figure of merit, $\Delta J_{cat}/J_0$, for $k_+ \rho_0=1$. Here we find that the auxiliary 
site can increase the reaction rate by a factor of $2.89$. This is a considerable effect, which is made possible due to the small value of $J_0$.

In the blocking regime the model exhibits a low tendency of molecules to stay bonded to the catalytic site and react. 
This follows from the transition rates $r_-=100 \gg r_+$,  meaning that
fuel molecules do not bind well to the catalytic site. (They have lower free energy in the bridging environment.) Moreover, the acceleration is maximal for
$r_-^* =100> k_r =10$. These rates mean that a fuel molecule in the active site is much more likely to return to the  bridging site than undergo the chemical reaction. 
Importantly, this probable scenario is only possible if the bridging site is empty, as fuel molecules block each other in our model. 

The mechanism utilizes this aspect of the system, and aims to increase the likelihood that once a molecule enters the active site, another one is brought to the bridging site. In this configuration, the molecule inside the active site is blocked, and the only process available to it is the chemical reaction. The likelihood of leaving the active site before reacting is significantly reduced since the rates for emptying the bridging site are smaller or roughly of the same order as $k_r$ (Specifically $\tilde{h}_+^*=1$ and $k_-^*$ between $1$ and $15$, depending on parameters). 

We turn to examine the processes involving the auxiliary site. When the active site is empty we have $h_+^*=100$ and $h_-^*=1$. These rates mean that a molecule in the bridging site is likely to be diverted to the auxiliary site and stay there. Once the catalytic site is occupied the rates change to $\tilde{h}_+^*=1$ and $\tilde{h}_-^*=100$. This causes the auxiliary site to release the fuel molecule it stored into the bridging site, which achieves the mechanism objective. Importantly, this process can compete with the escape 
from the active site. 

To conclude, in this regime the auxiliary site increases the current by blocking the escape of molecules from the catalytic site, thereby allowing the comparatively slow
reaction step to be more likely to occur. This conclusion is strengthened by examining the correlations between the occupations of the bridging site and the
catalytic site. For $k_+ \rho_0 = 10$ we find $\Pi (n_b=1 | n_c=1)= 0.887$, which can be compared to $\Pi (n_b=1 | n_c=0)= 0.567$, demonstrating that
the optimal choice of rates strongly favors filling the bridging site once the catalytic site is occupied.

\subsection{Molecule storage mechanism}

The second parameter regime is characterized by $r_+^*=100 > r_-^*$, which results in optimal auxiliary site rates of $h_-^*=\tilde{h}_+^*=100$ and $h_+^*=\tilde{h}_-^*=1$. Here the figure of merit $\Delta J_{cat}$ is again largest for moderate concentrations of fuel molecules. Quantitatively, the molecule storage mechanism has a more moderate effect on the reaction rate than the blocking mechanism. We find the largest effect for $k_+ \rho_0=10$, where $\Delta J_{cat}^* \simeq 0.738$, which should be compared to $J_0 \simeq 5.25$.

We again explain the mechanism by examining the values of the optimal rates. Here, typically $k_r > r_-^*$, suggesting that a molecule in the catalytic site is more likely to undergo the chemical reaction than to escape back to the bridging site. Furthermore, $r_+^* > r_-^*$ suggests that the catalytic site is highly effective in binding fuel molecules. In absence of the auxiliary site, a molecule in the bridging site has a probability of $r_+^*/(r_+^*+k_-^*)$ to bind to the active site, which is usually quite high. This picture is modified slightly when fuel molecules are in abundance, namely when $k_+ \rho_0=100$, where we find $k_-^*=100$. Using $k_-=1$ in such saturated conditions would have led to a near maximal value of $J_0$. The latter choice of rates is not likely to maximize $\Delta J_{cat}$, as it would have left very little room for improvement. The optimization solves this problem by finding parameters for which $J_0$ is not maximal.

To understand the role of the auxiliary site, we move on to analyze its optimal transition rates. When the active site is occupied we have $\tilde{h}_+^*=100$ and $\tilde{h}_-^*=1$. These values correspond to a setup that diverts a fuel molecule from the bridging site to the auxiliary site and store it there. Once the catalytic site is empty, the rates change
to $h_+^*=1$  and $h_-^*=100$, and the auxiliary site tends to push the molecule it (may have) stored back to the bridging site. Obviously, this chain of events can not occur if a fuel molecule moved from the catalytic site to the bridging site, due to blocking. It is only possible following a reaction.

All these considerations point out to a mechanism which we name the molecule storage mechanism. Here, the auxiliary site is used to store a fuel molecule when the active site is occupied and awaiting for a reaction. The 
molecule is then (likely to be) released to the bridging site once the reaction occurs. There, it is available to enter the catalytic site, decreasing the mean waiting time for the arrival of a new molecule. The effect of this mechanism can be seen by comparing the conditional probability $\Pi (n_a=1 | n_c=1 )=0.530$ to $\Pi (n_a=1 | n_c=0 )=0.0856$, so the auxiliary site is occupied mostly when the active site is.

\section{Studying the role of blocking in the bridging region}
\label{sec:nearby}

The model studied in previous sections assumed that the region near the active site functions like an effective site, which we termed the bridging  site.
We also assumed that this region holds a single fuel molecule at most, resulting in pronounced blocking effects.
So far we used this assumption because it led to a mathematically tractable model. However, it should be clear that this picture is an oversimplification.
Physically, fuel molecules are free to move around in the vicinity of the active site, and several molecules can be there simultaneously.

In this section, we relax this assumption on the bridging region by allowing it to hold multiple fuel molecules. Specifically, we allow
the occupation of the bridging region to take the values $n_b=0,1, \cdots, m$, meaning that blocking occurs only when the region is completely full, with $n_b=m$.
Based on our qualitative discussion of the mechanisms of acceleration, we expect that increasing the maximal occupancy will lead to a reduction in the efficacy 
of the blocking mechanism. Meanwhile, it should have a smaller effect on the current enhancement in the storage regime. In this section we test this by examining the increase in current for models with $m=2,3,4$.

The models we study have many-particle states that can be identified by the occupation numbers, $n_c, n_a, n_b$, as before. However, due to the larger range of $n_b$, each model has 
$4 \left(m+1\right)$ states. Rates that involve processes of {\em leaving} the bridging region must be modified to reflect the varying occupation. We make the simplifying assumption that the molecules move
freely in this region and explore it before they have a chance to leave. This suggests that the rate e.g. of entering the auxiliary site is of the form $\frac{n_m}{V} \hat{h}_+$, where
$V$ is the volume of the bridging region. To allow quantitative comparison with the results of Secs. \ref{sec:model}-\ref{sec:mechanism}, we set $h_+=\hat{h}_+/V$,
resulting in a transition rate of $n_b h_+$. The other rates of transitions in which a molecule leaves the bridging region, namely $\tilde{h}_+$, $r_+$, and $k_-$, are treated similarly. All other transitions have the same rates as their counterparts in the previous model.

We look for parameter regions where acceleration occurs for this family of models. The additional complexity due to the larger state space means that the optimization
for $\Delta J_{cat}$ and $\Delta J_{cat}/J_0$ is done numerically, using Matlab.
For each of the figures of merit we find two parameter regions with current enhancement. In both regions the optimal rates and behavior are similar to the ones found for the model studied in Secs.\ref{sec:model}-\ref{sec:mechanism}. 

\begin{figure}
	\centering
	\begin{subfigure}[b]{0.4\textwidth}
		\includegraphics[scale=0.2]{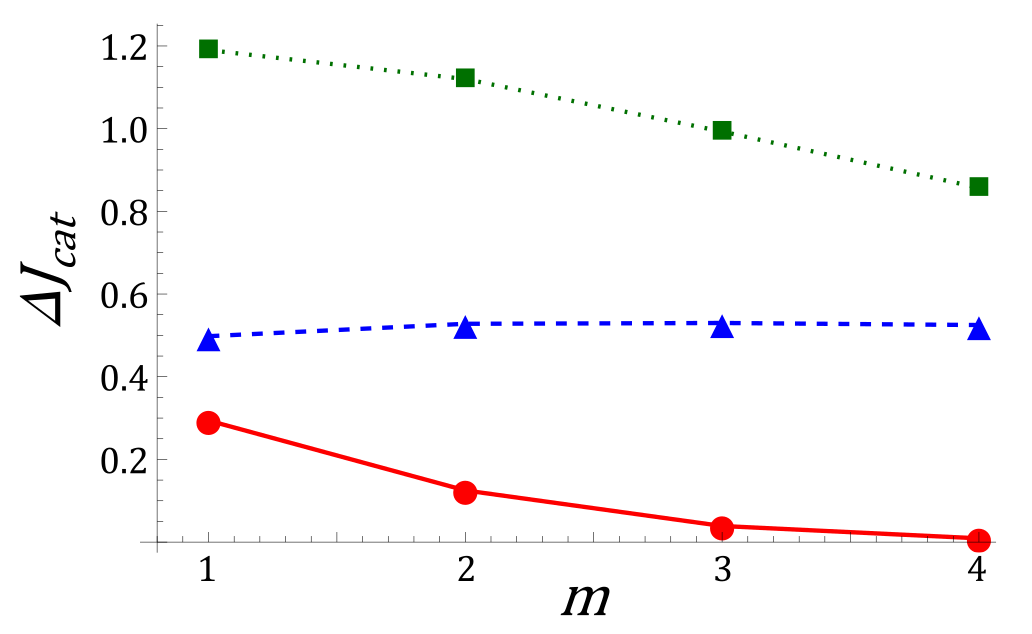}
		\caption{Blocking regime}
		\label{fig:delJblocking}
	\end{subfigure} 
	\begin{subfigure}[b]{0.4\textwidth}
		\includegraphics[scale=0.2]{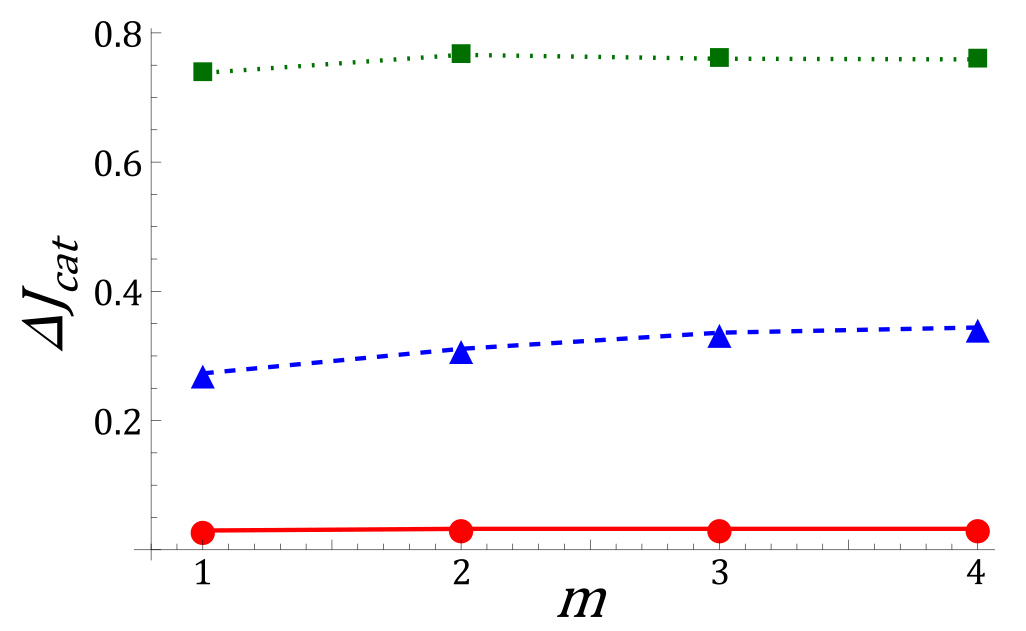}
		\caption{Molecule storage regime}
		\label{fig:delJstorage}
	\end{subfigure} 
	\begin{subfigure}[b]{0.4\textwidth}
	\includegraphics[scale=0.2]{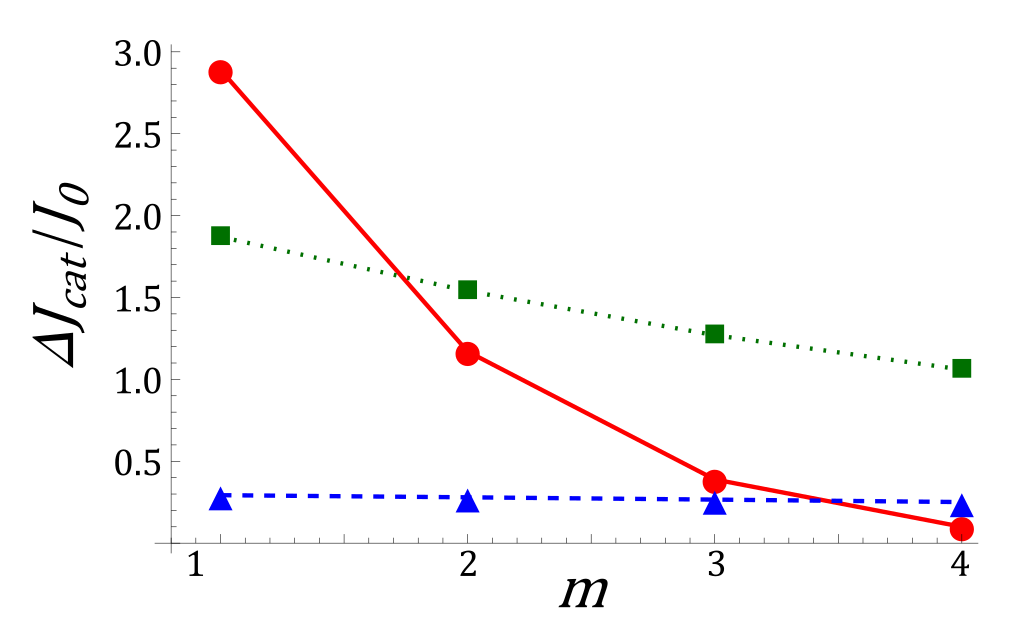}
	\caption{Blocking regime}
	\label{fig:ratioblocking}
\end{subfigure} 
\begin{subfigure}[b]{0.4\textwidth}
	\includegraphics[scale=0.2]{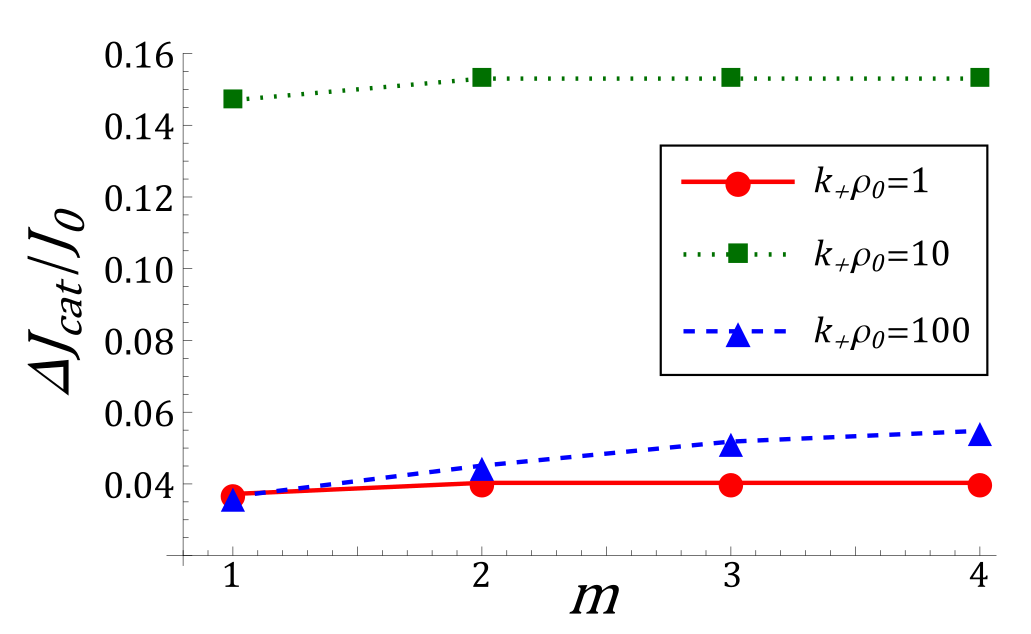}
	\caption{Molecule storage regime}
	\label{fig:ratiostorage}
\end{subfigure}   
	\caption{The figures of merit dependence on $m$, for different concentrations. The plot legends in panel (d) describe also the other plots. (a) shows $\Delta J_{cat}$ in the blocking regime, while (b) depicts $\Delta J_{cat}$ in molecule storage regime. (c) and (d) show $\Delta J_{cat}/J_0$ in the blocking and molecule storage regimes, respectively. The two figures of merit show qualitatively similar dependence on $m$.}
	\label{fig:graphm14}
\end{figure}

The results for the maximum of $\Delta J_{cat}$ are shown in Fig. \ref{fig:delJblocking} for the blocking regime, and Fig. \ref{fig:delJstorage} for the molecule storage regime. We can clearly see two different trends as a function of $m$. For $k_+ \rho_0=1,10$ in the blocking regime, $\Delta J_{cat}^*$ decreases significantly as $m$ increases. Meanwhile, for high concentration, $k_+ \rho_0=100$, $\Delta J_{cat}^*$ slightly increases when $m$ is increased from $1$ to $2$. $\Delta J^*_{cat}$  barely changes with further increase in $m$. Interestingly, the $m$-dependence of $\Delta J_{cat}^*$ in the storage regime for all values of $k_+ \rho_0$ is similar to that of the blocking regime for $k_+ \rho_0=100$. 
The results for $\left( \frac{\Delta J_{cat}}{J_0}\right)^*$ are shown in Figures \ref{fig:ratioblocking} and \ref{fig:ratiostorage}. They exhibit the same trends as the results for $\Delta J_{cat}^*$.

We identified the blocking and storage regimes by inspection of the optimal rates that result in maximal figures of merit. These rates are shown in Appendix \ref{app:oprates}.
In all of our results we found that two regimes can still be identified by using  Eqs. (\ref{eq:condblocking}) and (\ref{eq:condstorage}). Moreover, the rates are very similar to the optimal rates found in Sec. \ref{subsec:numopt}.
The main difference is that for $ k_+ \rho_0 =1$ in the molecule storage regime, we find $h_-^* \simeq 30$, which is no longer on the upper boundary of the allowed range. This reflect the fact that the argument given
in Sec. \ref{subsec:h} does not hold for models with $m>1$. We note that under these conditions, $\Delta J^*_{cat}$ depends weakly on $h_-$ in a wide range of values (roughly from 10 to 100), so that its value
at $h_-=100$ only differs by $1\times 10^{-4}$ from its maximal value.

The results for the family of models studied here reinforce our interpretation of the acceleration in the blocking regime as originating from
blocking the exit of the active site. Blocking only takes place when the bridging region is completely occupied, but this occurrence becomes less likely for larger values of $m$. 
This fits the decrease of both figures of merit as $m$ is increased (see Figures \ref{fig:delJblocking} and \ref{fig:ratioblocking}).
Interestingly, for large values of $k_+ \rho_0$, a different behavior is observed. Under such abundance of fuel molecule $\Delta J^*_{cat}$
barely changes for $m=2,3,4$. This behavior can be rationalized by noting that transitions towards the active site are slow. Due to this bottleneck the bridging region is likely to be completely full. Indeed, for $k_+ \rho_0=100$, we find that $0.45 \le \Pi (0 1 m) < 0.48$. This state, with a completely full bridging region, is therefore the most likely state of the system.

One can understand the persistence of current acceleration through blocking for $k_+ \rho_0=100$  by assuming that the system is at $(0 1 m)$ and considering the following sequence of events. To generate a current, a fuel molecule must enter the active site. While this step is unlikely, the same can be said about the same transition in the reference model. Once the system is at state $(11m-1)$, where $n_b=m-1$, there is a competition between several processes. One is a reaction at the catalytic site, but it is not as fast as other processes. The most rapid processes are escape from the active site, entry of an additional fuel molecule to the system, or release of a molecule from the auxiliary site. The last two processes fill the bridging region, thereby preventing the escape of the molecule from the active site.

These considerations suggest that after filling the catalytic site there is a probability roughly equal to $1/3$ that the molecule escape, and to $2/3$ that blocking occurs. 
These estimated probabilities lead to a configuration in which $k_r$ is one of the fastest rates. A similar argument can be made for the reference model without the auxiliary site. However, there the probability of 
blocking is proportional to $1/2$. This is the origin of the $m$-independence of $\Delta J^*_{cat}$ seen in Figure \ref{fig:delJblocking}.

The results for the storage regime mostly follow the expectation that increasing $m$ should not affect current enhancement. A single exception exists when $m$ is increased from $1$ to $2$. Then, a small jump in both figures of merit is seen. A possible explanation for this increase is that the storage mechanism can now operate also when there are some molecules in the bridging region. Specifically, a bridging region with a single molecule no longer blocks molecules from leaving the {\em auxiliary} site. 

\section{Conclusions}
\label{sec:conc}

In this work, we investigated the kinetic role of auxiliary binding sites. The term refers to sites that can bind and release molecules, but do not catalyze chemical reactions. Our goal was to discover if such sites can accelerate the reaction in a nearby catalytic site. We showed that such acceleration is possible if there are correlations between the dynamics of the auxiliary site and the state of the system. Importantly, we placed physically motivated restrictions on the auxiliary sites. They were meant to prevent shortcuts in which the auxiliary site creates an alternative path between
the molecule reservoir and the active site.

Our approach was to compare the reaction rate of simple models to almost-identical reference models without the auxiliary site. We found the transition rates that maximize the acceleration. This has led to identification of two distinct mechanisms of acceleration, which exist in different parameter regimes. For active sites that strongly bind fuel molecules, the increase in current is achieved through a storage and release mechanism. Here, the auxiliary site tends to store molecules when the active site is occupied, and
release them when the active site is empty. For poor binding sites, the increase in reaction rate is due to blocking. Here, the auxiliary site releases stored molecules in an attempt of delaying the escape of yet-to-react molecules from the catalytic site.

It is interesting to note that the auxiliary site plays a role similar to that of an 
autonomous Maxwell demon in bipartite jump processes \cite{Hartich_2014,PhysRevX.4.031015}. While our model is not 
bipartite, the auxiliary site effectively measures the state of the active site and responds to it by changing its rates between $h_\pm$ and $\tilde{h}_\pm$.
Interestingly, Eq. (\ref{eq:deltajc}) suggest that $\Delta J_{cat} \propto e^{\cal A}-Q$,
where ${\cal A}=\ln \frac{h_+ \tilde{h}_-}{\tilde{h}_+ h_-}$ is the thermodynamic affinity of the bold cycle in Fig. \ref{fig:8stscoopgraph}.
Also, $Q$ can be given information theoretic interpretation by recasting it as 
	\begin{equation}
	Q= \frac{\bar{\Pi}(n_b=1|n_c=1) \bar{\Pi}(n_b=0|n_c=0)}{\bar{\Pi}(n_b=0|n_c=1) \bar{\Pi}(n_b=1|n_c=0)}.
	\label{eq:qcond}
	\end{equation}
Here $\bar{\Pi} (n_b|n_c) \equiv \frac{\bar{\Pi} (n_c n_b)}{\sum_b \bar{\Pi} (n_c n_b) }$ is the steady-state conditional probability in the reference system.
The current acceleration is therefore proportional to a difference between a thermodynamic affinity and an information theoretic quantity.
While this is reminiscent of expressions appearing for Maxwell demons, the results of \cite{Hartich_2014,PhysRevX.4.031015} are not directly applicable for the current problem due to: i) the completely irreversible nature of the chemical reaction; ii) the non-bipartite structure; and iii) the focus on figures of merit that compare two different models.  

So far we examined the acceleration mechanisms as intriguing nonequilibrium phenomena. It is worthwhile to speculate on their relevance. The molecule storage mechanism seems to be biologically plausible. It requires catalytic sites that are effective binders. This matches the results for RecBCD \cite{zananiri2022auxiliary}, where it was shown that ATP binds preferably to the catalytic sites. For the blocking mechanism, a considerable increase can be found. However, this only occurs when the active site is not effective in capturing and keeping molecules, and the bridging region allows for blocking. Both assumptions are unlikely to hold for molecular motors. 

While the molecule storage mechanism may be biologically plausible, it would be a stretch to argue that it can fully explain the results of a specific biological motor such as the RecBCD studied in \cite{zananiri2022auxiliary}. To make such a claim, one would need direct evidence of coupling between the auxiliary and catalytic sites. In addition, the magnitude of acceleration is rather modest, reaching up to 15\%. Nevertheless, we expect that larger accelerations can be achieved by including additional effects that were not included in the models studied here. For instance, each catalytic site may have several adjacent auxiliary sites. Alternatively, one may include a mechanism where physical deformation of the protein changes the distance between
the auxiliary and active sites. This then leads to a shorter diffusion time from the auxiliary to the active site, increasing the reaction rate even further. Our results, therefore, suggest an interesting mechanism for accelerating catalytic reactions. But more detailed research, which takes into account system specific information, including the spatial arrangement of sites, is needed to determine if this mechanism contributes to specific biological processes.

\begin{acknowledgments}
We are grateful for support from the the Israel Science
Foundation (Grant No. 1929/21).
\end{acknowledgments}

\appendix
\section{The rate matrix}
\label{app:Rmat}
The explicit expression for the rate matrix appearing in Eq. (\ref{eq:mastereq}) is
\begin{equation*}
R=
\begin{pmatrix}
	R_{11} & k_- & k_r & 0 & 0 & 0 & 0 & 0\\
	k_+\rho_{0} &R_{22} & r_- & k_r & h_- & 0 & 0 & 0\\
	0 & r_+ & R_{33} & k_- & 0 & 0 & 0 & 0\\
	0 & 0 & k_+\rho_{0} & R_{44} & 0 & 0 & \tilde{h}_- & 0\\
	0 & h_+ & 0 & 0 & R_{55} & k_- & k_r & 0\\
	0 & 0 & 0 & 0 & k_+\rho_{0} & R_{66} & r_- & k_r\\
	0 & 0 & 0 & \tilde{h}_+ & 0 & r_+ & R_{77} & k_-\\
	0 & 0 & 0 & 0 & 0 & 0 & k_+\rho_{0} & R_{88}
\end{pmatrix}.
\end{equation*}
The diagonal elements are given by $R_{11}=-k_+\rho_{0}$, $R_{22}=-(h_++k_-+r_+)$, $R_{33}= -(k_r+r_-+k_+\rho_{0})$, $R_{44}=-(\tilde{h}_++k_-+k_r)$, $R_{55}=-(h_-+k_+\rho_{0})$, $R_{66}=-(k_-+r_+)$, $R_{77}= -(\tilde{h}_-+k_r+r_-+k_+\rho_{0})$, and $R_{88}=-(k_-+k_r)$.

\section{Explicit expression for the coefficient appearing in Eqs. (\ref{eq:deltajc}) and (\ref{eq:derivdelta})}
\label{app:coeffdelJ}

The coefficients in the expression for $\Delta J_{cat}$ are:
\begin{fleqn}
\begin{equation}
Q =\frac{k_-}{k_-+k_r}+\frac{k_r r_{+}}{(k_-+k_r)(k_r+r_{-})+k_r k_{+}\rho_{0}},
\label{eq:factornumerator}
\end{equation} 
\end{fleqn}
and
\begin{fleqn}
	\begin{equation}
	B =\frac{C_1}{N},
	\end{equation} 
\end{fleqn}
where
\begin{fleqn}
	\begin{eqnarray}
	C_1  & = &  k_r \left( k_+\rho_0 \right)^2 (k_- + k_r) (k_- + k_+ \rho_0) d_1, \\
	{\cal N} & = & C_2h_{-}+C_3h_{+}+C_4 \tilde{h}_-+C_5 \tilde{h}_++C_6 h_{-} \tilde{h}_-+C_7h_{+} \tilde{h}_++C_8h_{+}\tilde{h}_-+C_9h_{-}\tilde{h}_+. \\
	C_2 & = & \left[d_1+(k_- + k_r)k_rr_+\right] d_2^2, \\ 
	C_3 & = & k_+\rho_0 d_1  d_2^2, \\ 
	C_4 & = & r_+ k_+\rho_0 (k_- + k_r)  d_2^2, \\
	C_5 & = & r_+ {(k_+\rho_0)}^2  d_2^2, \\
	C_6 & = & (k_- + k_r)(k_- + r_+)  d_2^2, \\
	C_7 & = & k_+\rho_0 \left(k_r + r_- + k_+\rho_0 \right) d_2^2,\\
	C_8 & = &
	k_+\rho_0 \left(k_- + k_r \right) d_2
	\left[d_2+r_-r_+\left(k_+\rho_0+k_r+k_- \right)+
	k_r \left(r_+k_+\rho_0+k_-r_-+k_-k_+\rho_0 \right)\right], \\
	C_9 & = & d_2\left\{d_2 \left[k_-(k_r+k_-)+k_rr_+\right]+
	k_-(k_-+k_r)\left[k_-(k_r+k_-)+k_rr_+\right]k_+\rho_0 \right.  \nonumber \\ & + & 
	\left. \left[k_-(k_r+r_+)^2+k_rr_-(k_-+k_r)\right] (k_+\rho_0)^2+
	(k_- + r_+) (k_r + r_+) (k_+\rho_0)^3\right\}.
	\end{eqnarray}
\end{fleqn}
Here we defined
\begin{equation*}
d_1 \equiv (k_- + k_r) (k_r + r_-)\! +\! k_r k_+\rho_0,
\end{equation*}
and
\begin{multline*}
d_2 \equiv (k_- + k_r)\left[ k_- (k_r + r_-) + k_r r_+\right] + \left[k_r (k_r + r_- + r_+) +\right. \\ \left.
k_- (2 k_r + r_- + r_+)\right] k_+\rho_0 + (k_r + r_+) {(k_+\rho_0)}^2.
\end{multline*}

The coefficient in $\frac{\partial \Delta J_{cat}}{\partial h_{+}}$ is
\begin{multline}
	{\cal M} = \frac{C_1}{{\cal N}^2} \left[  C_2h_{-}\tilde{h}_{-}+C_4(\tilde{h}_{-})^2  +  C_5 \tilde{h}_{+} \tilde{h}_{-}+C_6 h_{-}(\tilde{h}_{-})^2  \right.  \\ 
	 +  \left. QC_3 \tilde{h}_{+}h_{-}+QC_7( \tilde{h}_{+})^2h_{-} +(C_9+QC_8) \tilde{h}_{+}h_{-} \tilde{h}_{-} \right].
\end{multline}

\section{Values of the optimal transition rates}
\label{app:oprates}

In this appendix we present the optimal rates that maximize the figures of merit $\Delta J_{cat}^* $ and $\left( \frac{\Delta J_{cat}}{J_0} \right)^*$. These rates were obtained 
as part of the optimization whose results were given in
 Sec. \ref{sec:max} and \ref{sec:nearby}. The models studied had different maximal occupation of the bridging region, namely $m=1-4$, as well as $k_+ \rho_0=1,10,100$. The values for $m=1$ are also included to allow for an easier comparison with the rates of higher $m$. The optimal rates for $\Delta J_{cat}^*$ are summarized in Tab. \ref{table:9} for the molecule storage regime, and Tab. \ref{table:10} for the blocking regime. Similarly, the rates of the optimal $\left( \Delta J_{cat}/J_0 \right)^*$ are presented in Tabs. \ref{table:11} and \ref{table:12} for the molecule storage and blocking regimes, respectively. The results show that the two mechanism identified for $m=1$ exist also for models with $m=2,3,4$. Namely, in all these cases the increased rate is either due to storage or blocking.

\begin{table}[h]
	\centering
	\begin{tabular}{|| c | c | c | c | c | c | c | c | c ||} 
		\hline
		$m$ & $k_+ \rho_0$ & $ r_+^*$ & $ r_-^*$ & $k_-^*$ & $h^*_-$ & $h^{*}_{+}$ & $\tilde{h}^{*}_{-}$ & $\tilde{h}^{*}_{+}$
		\\ [0.5ex] 
		\hline\hline
		\multirow{3}{10pt}{\centering 1} & 1 & 100 & 1 & 13.5 & 100 & 1 & 1 & 100\\ 
		& 10 & 100 & 1 & 11.3 & 100 & 1 & 1 & 100 \\
		& 100 & 100 & 3.22 & 100 & 100 & 1 & 1 & 100\\ [1ex] 
		\hline
		\multirow{3}{10pt}{\centering 2} & 1 & 100 & 1 & 13.8 & 29.4 & 1 & 1 & 100\\ 
		& 10 & 100 & 1 & 15.7 & 100 & 1 & 1 & 100 \\
		& 100 & 100 & 18.1 & 100 & 100 & 1 & 1 & 100\\ [1ex] 
		\hline
		\multirow{3}{10pt}{\centering 3} & 1 & 100 & 1 & 13.8 & 29.7 & 1 & 1 & 100\\ 
		& 10 & 100 & 1 & 16.8 & 100 & 1 & 1 & 100 \\
		& 100 & 100 & 24.4 & 100 & 100 & 1 & 1 & 100\\ [1ex] 
		\hline
		\multirow{3}{10pt}{\centering 4} & 1 & 100 & 1 & 13.8 & 29.8 & 1 & 1 & 100\\ 
		& 10 & 100 & 1 & 17.0 & 100 & 1 & 1 & 100 \\
		& 100 & 100 & 26.3 & 100 & 100 & 1 & 1 & 100\\ [1ex] 
		\hline
	\end{tabular}
	\caption{The transition rates that maximize $\Delta J_{cat}$ in the molecule storage regime.}
	\label{table:9}
\end{table}

\begin{table}[h]
	\centering
	\begin{tabular}{|| c | c | c | c | c | c | c | c | c ||} 
		\hline
		$m$ & $k_+ \rho_0$ & $ r_+^*$ & $ r_-^*$ & $k_-^*$ & $h^*_-$ & $h^{*}_{+}$ & $\tilde{h}^{*}_{-}$ & $\tilde{h}^{*}_{+}$
		\\ [0.5ex] 
		\hline\hline
		\multirow{3}{10pt}{\centering 1} & 1 & 6.87 & 100 & 1 & 1 & 100 & 100 & 1 \\ 
		& 10 & 14.4 & 100 & 1 & 1 & 100 & 100 & 1 \\
		& 100 & 13.3 & 100 & 3.93 & 1 & 100 & 100 & 1 \\ [1ex] 
		\hline
		\multirow{3}{10pt}{\centering 2} & 1 & 3.60 & 100 & 1 & 1 & 100 & 100 & 1 \\ 
		& 10 & 8.41 & 100 & 1 & 1 & 100 & 100 & 1 \\
		& 100 & 8.27 & 100 & 3.73 & 1 & 100 & 100 & 1 \\ [1ex] 
		\hline
		\multirow{3}{10pt}{\centering 3} & 1 & 2.40 & 100 & 1 & 1 & 100 & 100 & 1 \\ 
		& 10 & 6.08 & 100 & 1 & 1 & 100 & 100 & 1 \\
		& 100 & 6.13 & 100 & 3.20 & 1 & 100 & 100 & 1 \\ [1ex] 
		\hline
		\multirow{3}{10pt}{\centering 4} & 1 & 1.66 & 100 & 1 & 1 & 100 & 100 & 1 \\ 
		& 10 & 4.79 & 100 & 1 & 1 & 100 & 100 & 1 \\
		& 100 & 4.92 & 100 & 2.80 & 1 & 100 & 100 & 1 \\ [1ex] 
		\hline
	\end{tabular}
	\caption{The transition rates that maximize $\Delta J_{cat}$ in the blocking regime.}
	\label{table:10}
\end{table}

\begin{table}[h]
	\centering
	\begin{tabular}{|| c | c | c | c | c | c | c | c | c ||} 
		\hline
		$m$ & $ k_+ \rho_0$ & $ r_+^*$ & $ r_-^*$ & $k_-^*$ & $h^*_-$ & $h^{*}_{+}$ & $\tilde{h}^{*}_{-}$ & $\tilde{h}^{*}_{+}$
		\\ [0.5ex] 
		\hline\hline
	\multirow{3}{10pt}{\centering 1} & 1 & 100 & 1 & 18.8 & 100 & 1 & 1 & 100\\ 
		& 10 & 100 & 1 & 21.8 & 100 & 1 & 1 & 100\\
		& 100 & 100 & 13.5 & 100 & 100 & 1 & 1 & 100\\ [1ex] 
		\hline
	\multirow{3}{10pt}{\centering 2} & 1 & 100 & 1 & 18.8 & 28.2 & 1 & 1 & 100\\ 
	    & 10 & 100 & 1 & 26.8 & 100 & 1 & 1 & 100\\
		& 100 & 100 & 37.2 & 100 & 100 & 1 & 1 & 100\\ [1ex] 
		\hline
	\multirow{3}{10pt}{\centering 3} & 1 & 100 & 1 & 18.8 & 28.4 & 1 & 1 & 100\\ 
		& 10 & 100 & 1 & 27.6 & 100 & 1 & 1 & 100\\
		& 100 & 84.4 & 45.1 & 100 & 100 & 1 & 1 & 100\\ [1ex] 
		\hline
	\multirow{3}{10pt}{\centering 4} & 1 & 100 & 1 & 18.8 & 28.2 & 1 & 1 & 100\\ 
    	& 10 & 100 & 1 & 27.7 & 100 & 1 & 1 & 100\\
	    & 100 & 68.8 & 43.2 & 100 & 100 & 1 & 1 & 100\\ [1ex] 
	    \hline
	\end{tabular}
	\caption{The transition rates for maximal $\Delta J_{cat}/J_0$ in the molecule storage regime.}
	\label{table:11}
\end{table}

\begin{table}[h]
	\centering
	\begin{tabular}{|| c | c | c | c | c | c | c | c | c ||} 
		\hline
		$m$ & $k_+ \rho_0$ & $ r_+^*$ & $ r_-^*$ & $k_-^*$ & $h^*_-$ & $h^{*}_{+}$ & $\tilde{h}^{*}_{-}$ & $\tilde{h}^{*}_{+}$
		\\ [0.5ex] 
		\hline\hline
		\multirow{3}{10pt}{\centering 1} & 1 &1 & 100 & 1.25 & 1 & 100 & 100 & 1\\ 
		& 10 & 1 & 100 & 1 &  1 & 100 & 100 & 1\\
		& 100 & 1 & 100 & 15.9 & 1 & 100 & 100 & 1\\ [1ex] 
		\hline
		\multirow{3}{10pt}{\centering 2} & 1 &1 & 100 & 1 & 1 & 100 & 100 & 1\\ 
		& 10 & 1 & 100 & 1 &  1 & 100 & 100 & 1 \\
		& 100 & 1 & 100 & 10.3 & 1 & 100 & 100 & 1\\ [1ex] 
		\hline
		\multirow{3}{10pt}{\centering 3} & 1 &1 & 100 & 1 & 1 & 100 & 100 & 1\\ 
		& 10 & 1 & 100 & 1 &  1 & 100 & 100 & 1 \\
		& 100 & 1 & 100 & 7.78 & 1 & 100 & 100 & 1\\ [1ex] 
		\hline
		\multirow{3}{10pt}{\centering 4} & 1 &1 & 100 & 1 & 1 & 100 & 100 & 1\\ 
		& 10 & 1 & 100 & 1 &  1 & 100 & 100 & 1\\
		& 100 & 1 & 100 & 6.35 & 1 & 100 & 100 & 1\\ [1ex] 
		\hline
	\end{tabular}
	\caption{The transition rates that maximize $\Delta J_{cat}/J_0$ in the blocking regime.}
	\label{table:12}
\end{table}
\bibliography{refauxiliary}
\end{document}